\documentclass[aps,prb,amsmath,amssymb,reprint,superscriptaddress]{revtex4-2}

\usepackage{graphicx}
\usepackage{amsmath,amssymb}
\usepackage{xcolor}
\usepackage{color,soul}
\begin{document}

\author{P. L. Rodr\'iguez-Kessler}
\email{plkessler@cio.mx}
\affiliation{Centro de Investigaciones en \'Optica A.C., Loma del Bosque 115, Lomas del Campestre, Leon, 37150, Guanajuato, Mexico} 

%\title{Triangle-supported-wheel Geometry of the B$_8$Cu$_3^{-}$ Cluster: A DFT Study}
\title{Structural prediction of B$_{18}$Y$_{2}$ cluster: A Machine-Learning-Assisted Basin-Hopping Study}

\date{\today}

\begin{abstract}
The structural and optical properties of the doubly yttrium-doped boron cluster B$_{18}$Y$_2$ have been systematically investigated using density functional theory calculations. The lowest-energy structure was identified through extensive basin-hopping searches accelerated by a pre-trained MACE machine-learning potential and subsequently refined and validated at the DFT level. The resulting global-minimum structure adopts a double-ring geometry, consisting of two fused B$_9$ rings stabilized by yttrium atoms positioned above and below the boron framework. Vibrational frequency calculations confirm the dynamical stability of the optimized structure, while the calculated infrared and UV--Vis spectra provide characteristic signatures of the Y--B interactions and the electronic structure of the boron framework. These results demonstrate how yttrium doping can stabilize unusual double-ring boron architectures and highlight the effectiveness of machine-learning-assisted basin-hopping searches for exploring the complex potential-energy landscapes of doped boron clusters.
\end{abstract}

\maketitle

\section{Introduction}

Boron clusters have emerged as versatile platforms in cluster science owing to their remarkable structural diversity and their distinctive electron-deficient bonding, which frequently involves multicenter and highly delocalized interactions.\cite{C6CC09570D,C9CP03496J} In particular, medium-sized boron clusters can develop extended ring-like architectures in which conventional two-center bonding gives way to collective bonding patterns. The B$_{18}$ cluster is especially interesting in this regard, as its low-energy structures have been associated with double-ring and toroidal frameworks that provide large delocalized boron networks.\cite{doi:10.1021/jp8087918,C6SC02623K} Introducing metal atoms into such frameworks offers an additional means of controlling their geometry and electronic structure, giving rise to unusual bonding motifs and architectures such as inverse-sandwich configurations,\cite{JIA2014128,Zhuan-Yu2014,PHAM2019186,C5CP01650A,LI202325821,RODRIGUEZKESSLER2025117486,b7al2,b7cr2} metal-centered aromaticity, and extended toroidal structures with substantial electron delocalization.\cite{doi:10.1021/acs.inorgchem.7b02585,doi:10.1021/acs.jpclett.0c02656}

Recent studies from our group have extended the inverse-sandwich concept to the B$_{18}$ framework, demonstrating that double-ring motifs can be stabilized by different metal centers. In particular, our investigations of B$_{18}$Zr$_2$, B$_{18}$Ca$_2$, and B$_{18}$Ag$_2$ revealed closely related double-ring architectures in which two metal atoms interact cooperatively with the extended boron framework.\cite{RKessler2026,RodriguezKessler2026_JCC,b18ag2} These results indicate that the inverse-sandwich motif is not restricted to the smaller boron rings commonly considered in metal-doped boron clusters, but can be extended to larger boron frameworks containing two fused rings. At the same time, the nature of the metal centers can substantially influence the relative stability, electronic structure, and bonding characteristics of these systems. This emerging family of B$_{18}$M$_2$ clusters therefore provides a useful framework for examining how different metals modulate an extended boron scaffold while preserving the characteristic double-ring topology.

Yttrium is particularly attractive in this context because of its electropositive character, accessible valence orbitals, and ability to participate in bonding with electron-deficient boron frameworks. Its position between the alkaline-earth and transition-metal regions of the periodic table also provides an opportunity to examine how the electronic characteristics of the metal influence the stability of the B$_{18}$ double-ring architecture. This question is particularly relevant in light of the unusual bonding encountered in smaller metal-doped boron clusters, such as B$_7$M$_2$ systems, in which metal atoms interact cooperatively with a boron ring and contribute to extended $\sigma$- and $\pi$-bonding networks.\cite{D5CP01078K,https://doi.org/10.1002/adts.70446,ALYASSIRI2026116025,GUEVARAVELA2025115487,10.1039/d4cp04444d,RODRIGUEZKESSLER2025117486,OLALDELOPEZ2025419} Extending this concept to B$_{18}$Y$_2$ allows the structural and electronic consequences of Y incorporation to be examined within the broader B$_{18}$M$_2$ family and provides an opportunity to assess whether the double-ring inverse-sandwich motif remains favorable for a different metal center.

In the present work, we investigate the B$_{18}$Y$_2$ cluster by combining machine-learning-assisted global structure optimization with density functional theory calculations. The global search was performed using an adaptive basin-hopping strategy accelerated by a pre-trained MACE potential, followed by DFT refinement of the lowest-energy structures identified during the search. The resulting low-energy configurations were analyzed in terms of their relative energies, spin states, vibrational stability, and spectroscopic properties. Particular attention is given to the structural characteristics of the lowest-energy double-ring configuration and its infrared and UV--Vis signatures. This study extends our recent investigations of B$_{18}$M$_2$ clusters and provides further insight into the structural diversity and metal-dependent stabilization of double-ring boron architectures.

\section{Computational Details}

The potential-energy surface of B$_{18}$Y$_2$ was explored using an adaptive basin-hopping (ABH) scheme,\cite{Carmona2026} accelerated by a pre-trained MACE machine-learning potential. A total of 300 ABH steps were performed, with each trial structure locally optimized using MACE and the BFGS algorithm. The maximum atomic displacement was adaptively adjusted every 20 steps to maintain an acceptance ratio close to 50%, while a Metropolis criterion with $k*{\mathrm{B}}T=0.10$ eV was used to accept or reject trial configurations. The 50 lowest-energy structures obtained from the MACE-assisted search were selected for subsequent DFT refinement.

All DFT calculations were performed using ORCA 6.0.0.\cite{10.1063/5.0004608} Geometry optimizations were carried out using the PBE0 hybrid functional with the Def2-TZVP basis set,\cite{10.1063/1.478522,B508541A} including Grimme's DFT-D3(BJ) dispersion correction. A quasi-Newton optimization procedure based on the BFGS algorithm was employed with the TightSCF convergence criterion. The selected low-energy structures were optimized for spin multiplicities ranging from singlet to quintet, and the lowest-energy electronic state was subsequently identified. Harmonic vibrational frequency calculations were performed to verify the nature of the optimized structures as true local minima.

The electronic structure and bonding characteristics of the lowest-energy configuration were further examined using electron-density-based analyses. Electron localization function (ELF), charge-distribution, and related real-space analyses were performed using numerical grid data analyzed with Multiwfn.\cite{https://doi.org/10.1002/jcc.22885}

\section{Results and Discussion}

\subsection{Global Structure Search}

\begin{figure}[ht]
\centering
\includegraphics[width=0.45\textwidth]{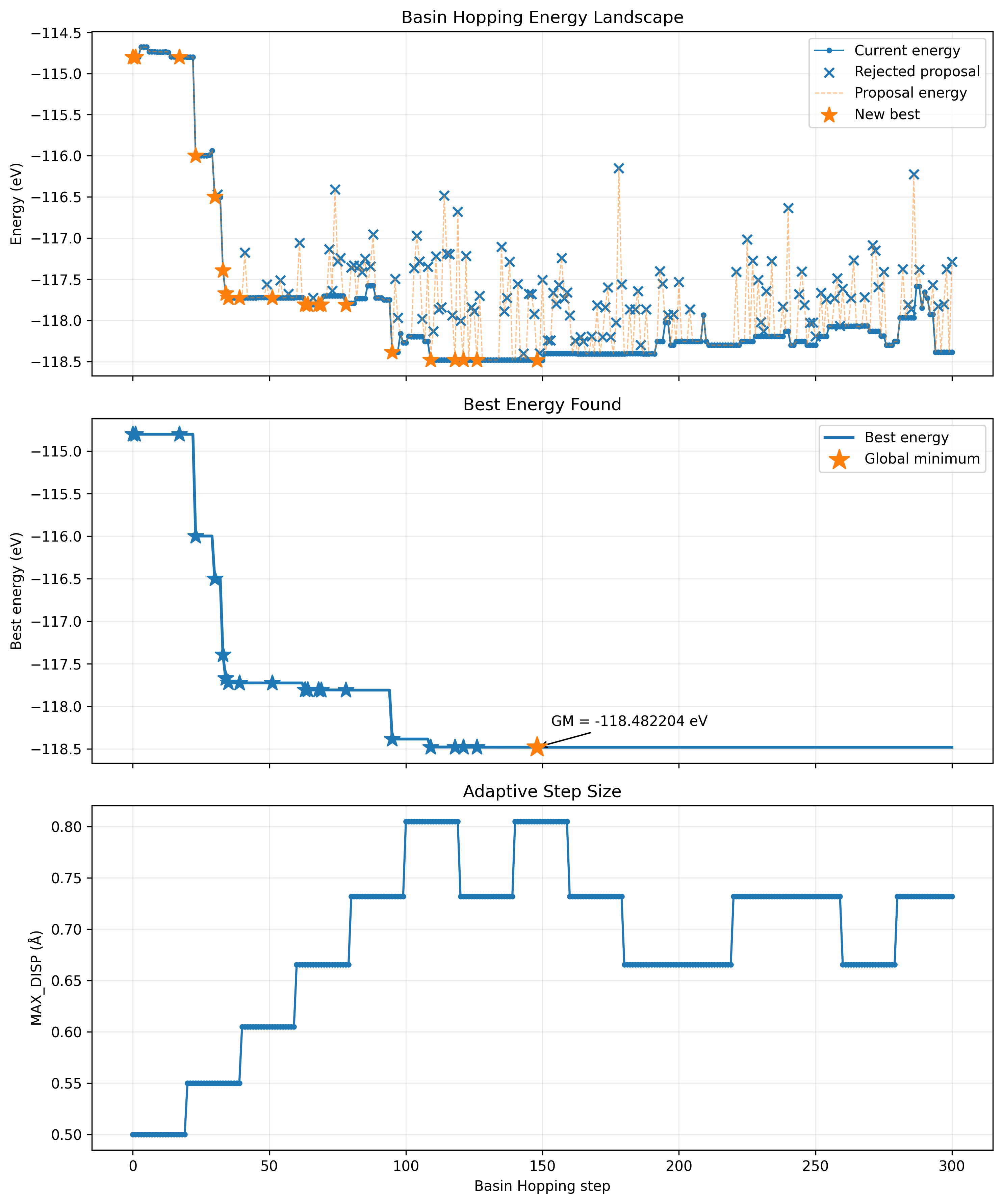}
\caption{Potential-energy landscape obtained from the 300-step MACE-assisted adaptive basin-hopping search of B$_{18}$Y$_2$. The upper panel shows the energies of accepted and rejected configurations, the middle panel displays the lowest energy found during the search, and the lower panel shows the evolution of the adaptive maximum displacement.}
\label{bh:curves}
\end{figure}

The structural exploration of the B$_{18}$Y$_2$ cluster was initially performed using the adaptive basin-hopping (ABH) approach combined with the pre-trained MACE machine-learning potential. A total of 300 basin-hopping steps were carried out to explore the low-energy region of the potential-energy surface. As shown in Figure~\ref{bh:curves}, the search rapidly identified progressively lower-energy configurations during the initial stages, followed by an extensive exploration of a relatively narrow low-energy region. The energy of the initial relaxed structure was approximately $-114.8$ eV, whereas successive basin-hopping moves led to substantially lower-energy configurations. Several pronounced decreases in the energy landscape are observed during the first $\sim100$ steps, indicating the discovery of distinct structural basins.

After approximately 100 steps, the search entered a low-energy region centered around $-118.5$ eV, where numerous structurally different configurations were sampled. The repeated transitions between nearby minima, together with occasional uphill moves accepted through the Metropolis criterion, demonstrate that the ABH procedure continued to explore the potential-energy surface rather than becoming trapped in a single minimum. The lowest-energy structure identified during the MACE-assisted search has an energy of $-118.482204$ eV and was reached at approximately step 140. Importantly, this configuration remained the lowest-energy structure throughout the remainder of the 300-step search, providing a robust candidate for the global minimum.

The adaptive displacement scheme also contributed to maintaining efficient exploration of the potential-energy surface. Starting from a maximum displacement of 0.50~\AA, the displacement amplitude increased progressively as the search proceeded, reaching values of approximately 0.80~\AA. Subsequent adjustments maintained the displacement within the range of approximately 0.65--0.80~\AA, reflecting the response of the algorithm to the observed acceptance ratio. This behavior indicates that the adaptive strategy was able to maintain substantial structural exploration while avoiding excessively large perturbations that could lead to inefficient sampling.

The MACE-assisted ABH search therefore provided an efficient first-stage exploration of the B$_{18}$Y$_2$ potential-energy surface and identified a well-defined low-energy candidate. To account for the limitations of the machine-learning potential and to establish the energetic ordering at a higher level of theory, the 50 lowest-energy structures obtained from the search were subsequently subjected to DFT geometry optimization for spin multiplicities ranging from singlet to quintet. The resulting DFT structures and relative energies are discussed below.

\begin{figure}[ht]
\centering
\includegraphics[width=0.45\textwidth]{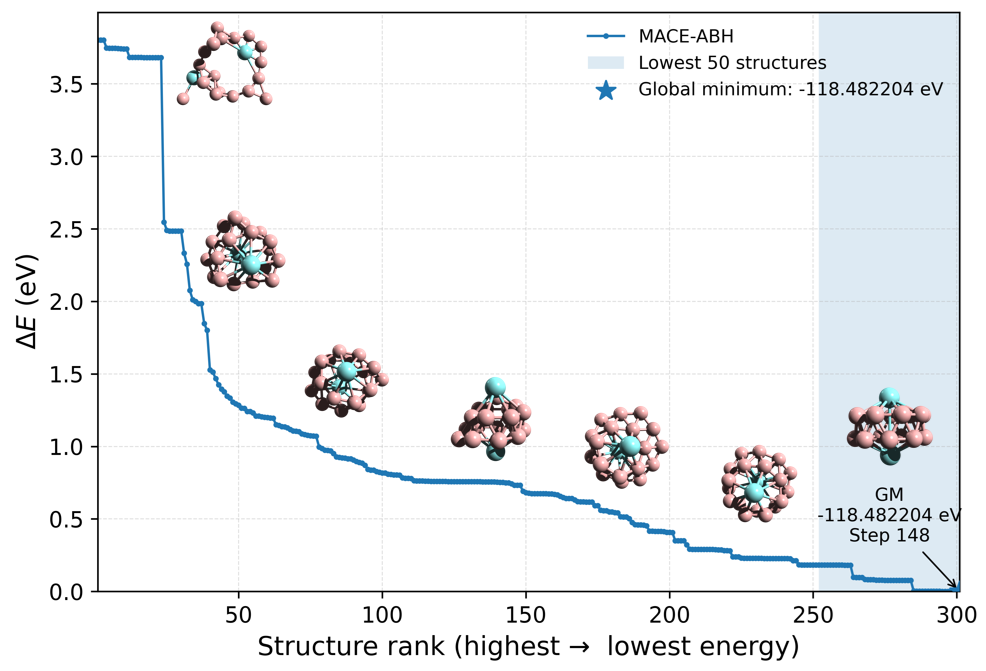}
\caption{Energy ranking of the lowest-energy structures obtained during the ABH search using the MACE potential. Representative structures along the energy landscape are shown.}
\label{bh:high2low}
\end{figure}

Next, we analyze the energetic ranking of the lowest-lying structures identified during the ABH search using the MACE potential. Figure~\ref{bh:high2low} shows the relative energies of the selected structures with respect to the global minimum, together with representative configurations along the energy landscape. The search rapidly explores and stabilizes the system during the initial stages, producing a pronounced decrease in energy as the Y atoms progressively rearrange with respect to the boron framework. At higher ranks, the energy decrease becomes more gradual and develops a step-like profile, reflecting the successive sampling of distinct local minima and their structural relaxation. The lowest-energy structures exhibit increasingly similar double-ring configurations, with the Y atoms adopting positions that stabilize the B$_{18}$ framework. The global minimum is identified at the lowest-energy end of the ranking, corresponding to the most stable arrangement obtained from the MACE-assisted ABH exploration.

\begin{figure}[ht]
\centering
\includegraphics[width=0.45\textwidth]{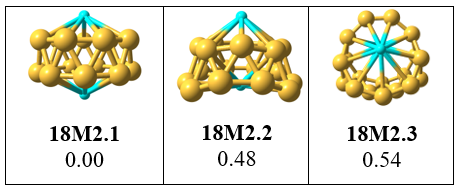}
\caption{Lowest energy structures for B$_{18}$Y$_2$ cluster at PBE0/def2-TZVP.}
\label{fig_geom}
\end{figure}

\subsection{Relative DFT energies}

The relative energies reveal a clear energetic preference for the double-ring structure, which is identified as the global minimum (GM) at the PBE0 level of theory. The bent double-ring and single-ring isomers lie 0.48 and 0.54~eV higher in energy, respectively, indicating that the double-ring motif is strongly favored energetically (Figure~\ref{fig_geom}). The three lowest-energy structures have spin multiplicities of 3, 3, and 1, respectively (Table~\ref{table1}). The same energetic ordering is also obtained with the $\omega$B97X-D3 functional, further supporting the robustness of the double-ring motif as the preferred structural arrangement for the B$_{18}$Y$_2$ cluster.

\begin{table}[h]
\centering
\setlength{\tabcolsep}{12pt} % default is 6pt
\caption{Spin multiplicities and relative energies (in eV) of lowest B$_{18}$Y$_2$ isomers at the PBE0 level. The global minimum is denoted by GM.}
\label{table1}
\begin{tabular}{lcc}
\hline\hline
Isomer & M & $\Delta{E}$ \\
\hline
Double-ring structure (GM) & 3 & 0.0  \\
Bent double-ring isomer & 3 & 0.48  \\
Single-ring structure & 1 & 0.54  \\
\hline\hline
\end{tabular}
\end{table}

\subsection{Vibrational and Optical Properties}

The vibrational and electronic absorption properties of the B$_{18}$Y$_2$ double-ring cluster were investigated to elucidate the relationship between its structure, bonding, and spectroscopic response. The calculated IR spectrum spans approximately 177--1361~cm$^{-1}$, and the absence of imaginary frequencies confirms the vibrational stability of the optimized structure.

The most intense IR bands occur below 450~cm$^{-1}$, particularly at 386.64 and 427.41~cm$^{-1}$, with intensities of 78.20 and 92.11~km~mol$^{-1}$, respectively. An additional band at 408.58~cm$^{-1}$ has an intensity of 24.13~km~mol$^{-1}$. These modes are associated with coupled Y--B motions and collective deformations of the boron framework, reflecting the interaction between the Y centers and the fused B$_9$ rings. Weaker low-frequency modes, including those at 188.90~cm$^{-1}$ (33.35~km~mol$^{-1}$) and 177.09~cm$^{-1}$ (6.20~km~mol$^{-1}$), further reflect collective motions involving the Y atoms and boron framework.

Between 450 and 900~cm$^{-1}$, the spectrum is dominated by weaker bands associated mainly with B--B skeletal deformations. At higher frequencies, the intensities remain generally low, except for the band at 1013.78~cm$^{-1}$ with an intensity of 15.60~km~mol$^{-1}$. These features are primarily related to localized B--B stretching vibrations. Overall, the IR spectrum provides a characteristic vibrational signature of the B$_{18}$Y$_2$ double-ring structure, with the strongest absorptions arising from low-frequency motions involving the Y--B framework.

\begin{figure*}[ht]
\centering
%\fbox{\parbox{0.45\textwidth}{
%}}
\begin{tabular}{cc}
 \resizebox*{0.40\textwidth}{!}{\includegraphics{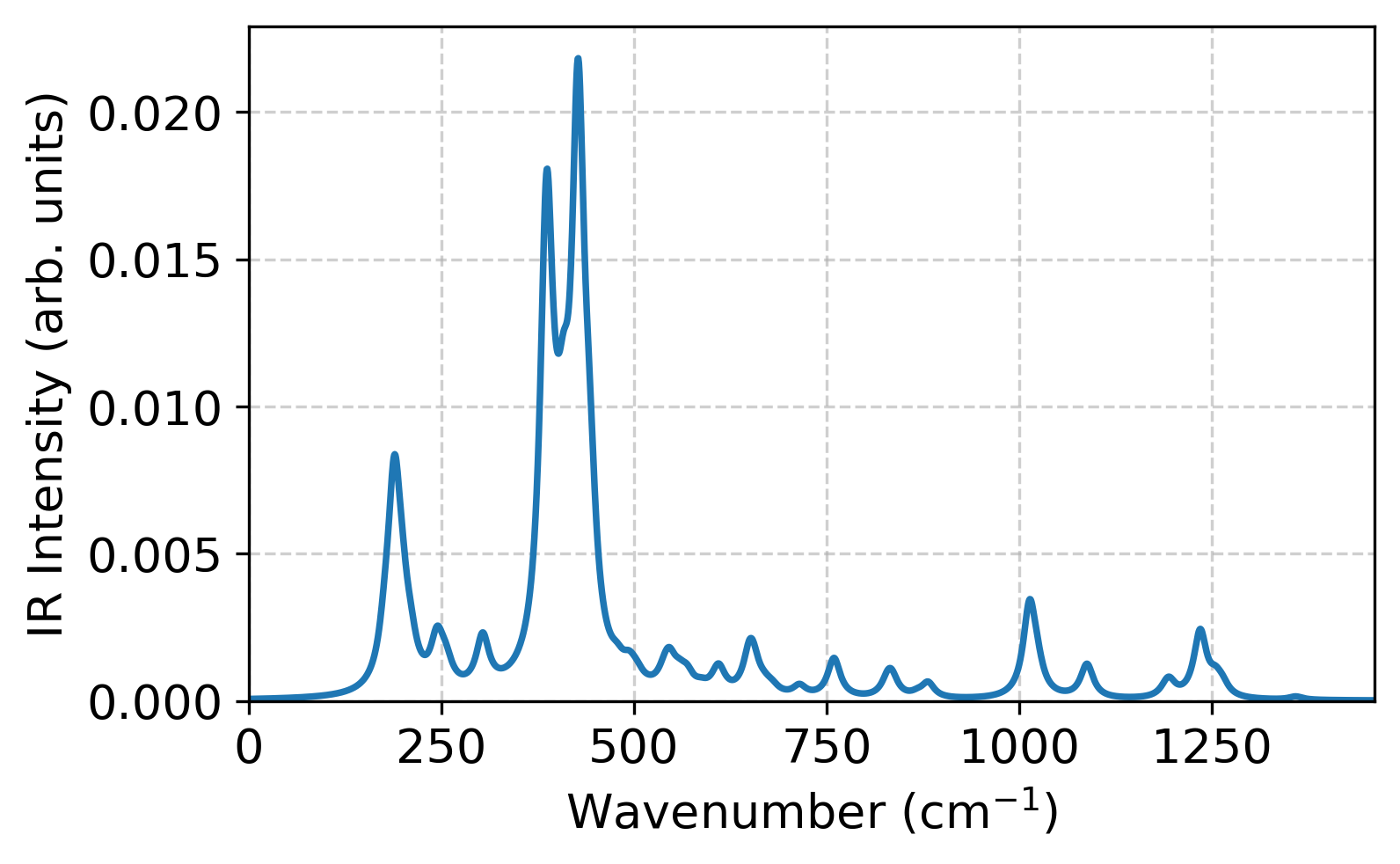}} &
\resizebox*{0.39\textwidth}{!}{\includegraphics{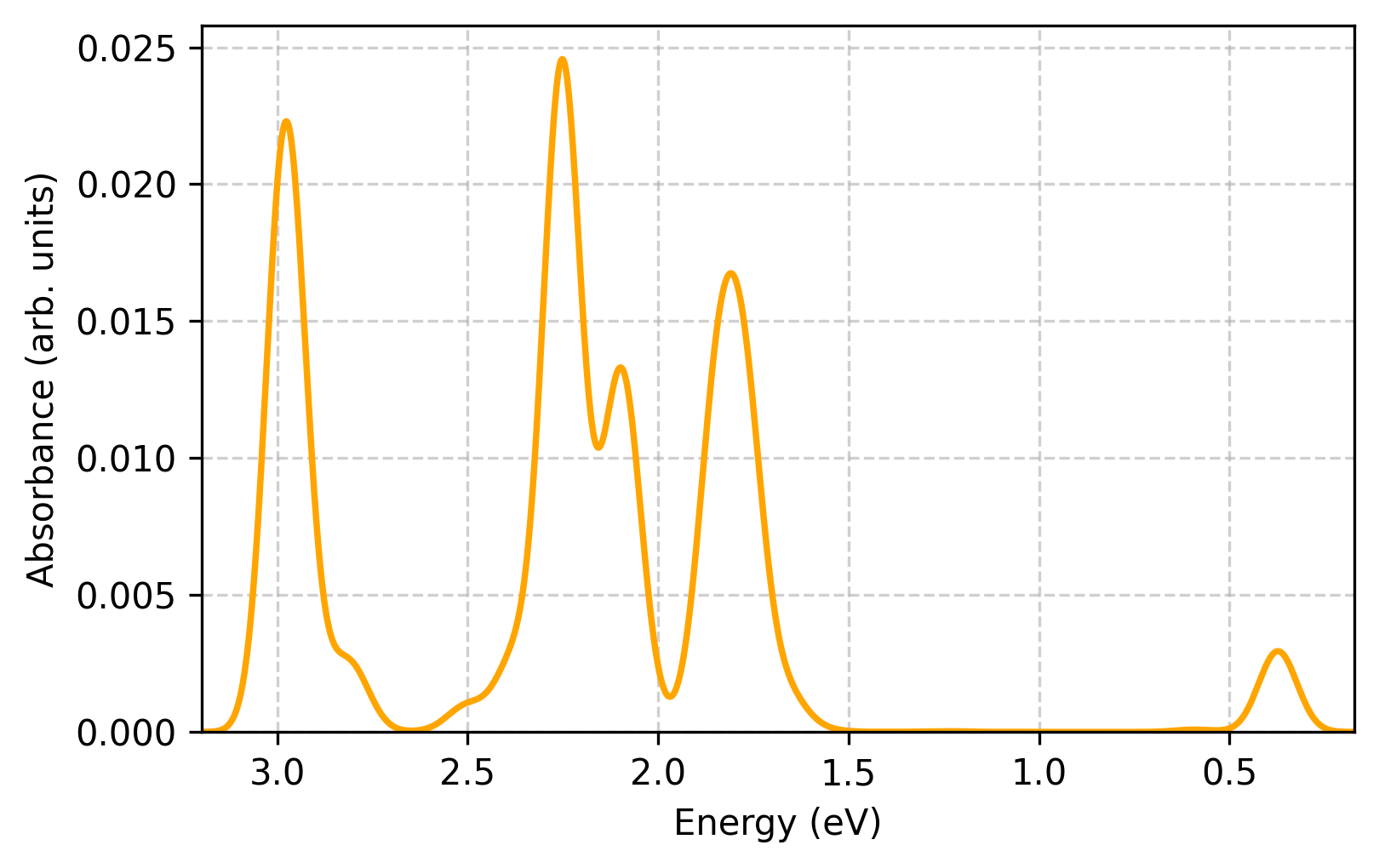}} \\
\end{tabular}
\caption{IR and UV-VIS spectra of B$_{18}$Y$_2$ cluster.}
\label{fig_elf_lol}
\end{figure*}

The calculated UV--Vis spectrum of B$_{18}$Y$_2$ exhibits a broad optical response extending from the near-infrared to the near-ultraviolet region. The lowest-energy transition occurs at 0.372~eV (3328.7~nm), with a relatively small oscillator strength of 0.00295, indicating a weak low-energy optical response. This feature is followed by a broad energy range from approximately 0.59 to 1.6~eV that is essentially dark, with no significant absorption features. More pronounced absorption features emerge above 1.7~eV, including transitions at 1.771 and 1.835~eV with oscillator strengths of 0.01046 and 0.00685, respectively.
\\
The strongest optical response is concentrated between approximately 2.1 and 3.0~eV. In particular, transitions at 2.251 and 2.252~eV exhibit oscillator strengths of 0.01194 and 0.01242, while the most intense transition occurs at 2.978~eV (416.3~nm), with an oscillator strength of 0.02027. Additional moderately intense transitions are observed at 2.086--2.128~eV and around 2.37~eV. These features indicate that the B$_{18}$Y$_2$ cluster possesses appreciable optical activity across the visible region. Overall, the calculated spectrum reveals a relatively weak low-energy response followed by a series of more intense electronic transitions in the visible and near-UV regions, providing a characteristic optical signature of the B$_{18}$Y$_2$ double-ring structure.

\section{Conclusions}

In summary, the B$_{18}$Y$_2$ cluster was investigated by combining machine-learning-assisted basin-hopping global optimization with DFT calculations. The MACE-assisted search efficiently explored the potential-energy surface and identified a low-energy double-ring structure consisting of two fused B$_9$ rings stabilized by Y atoms located above and below the boron framework. DFT calculations at the PBE0/def2-TZVP level identify this structure as the lowest-energy isomer, lying 0.48 and 0.54~eV below the bent double-ring and single-ring structures, respectively. The same energetic ordering obtained with $\omega$B97X-D3 supports the robustness of the double-ring motif. The absence of imaginary vibrational frequencies confirms the dynamical stability of the lowest-energy structure. Its calculated IR spectrum is dominated by low-frequency modes involving coupled Y--B and boron-framework motions, while the UV--Vis spectrum shows a series of electronic transitions extending from the near-infrared to the near-ultraviolet regions, with the strongest absorption features occurring in the visible range. Overall, these results demonstrate that yttrium doping can stabilize an extended double-ring boron architecture and significantly influence its vibrational and optical properties. The combination of machine-learning-assisted basin-hopping and DFT refinement provides an efficient approach for exploring metal-doped boron clusters and can be extended to related B$_{18}$M$_2$ systems.

%\clearpage

\section{Acknowledgments}
P.L.R.-K. would like to thank the support of Centro de Investigaciones en \'Optica A.C., CIO. 

%\bibliographystyle{ieeetr}

%%%%%%%%%%%%%%
% References %
%%%%%%%%%%%%%%

\bibliographystyle{unsrt}
\bibliography{mendelei.bib}
%\section{References}
%\nocite{*}
\end{document}